\documentclass[epjST]{svjour}
\usepackage{graphicx}
\usepackage{amsmath}
\usepackage{hyperref}
\begin{document}
\title{Constraint methods for determining pathways and free energy of activated processes}
\author{J\"{u}rgen Schlitter\thanks{\email{\href{mailto:juergen.schlitter@rub.de}{juergen.schlitter@rub.de}}}}
\institute{Ruhr-Universit\"{a}t Bochum ND04, 44780 Bochum, Germany}
\abstract{
Activated processes from chemical reactions up to conformational transitions of large biomolecules are hampered by barriers which are overcome only by the input of some free energy of activation. Hence, the characteristic and rate-determining barrier regions are not sufficiently sampled by usual simulation techniques. Constraints on a reaction coordinate \textit{r} have turned out to be a suitable means to explore difficult pathways without changing potential function, energy or temperature. For a dense sequence of values of \textit{r}, the corresponding sequence of simulations provides a pathway for the process. As only one coordinate among thousands is fixed during each simulation, the pathway essentially reflects the system's internal dynamics. From mean forces the free energy profile can be calculated to obtain reaction rates and insight in the reaction mechanism. In the last decade, theoretical tools and computing capacity have been developed to a degree where simulations give impressive qualitative insight in the processes at quantitative agreement with experiments.
Here, we give an introduction to reaction pathways and coordinates, and develop the theory of free energy as the potential of mean force. We clarify the connection between mean force and constraint force which is the central quantity evaluated, and discuss the mass metric tensor correction. Well-behaved coordinates without tensor correction are considered. We discuss the theoretical background and practical implementation on the example of the reaction coordinate of targeted molecular dynamics simulation. Finally, we compare applications of constraint methods and other techniques developed for the same purpose, and discuss the limits of the approach.
} 
\newpage
\maketitle
\newpage
\section{Introduction}
\label{intro}
In the last decade, theoretical tools and computing capacity have been developed to a degree where simulations give impressive qualitative insight in activated processes at quantitative agreement with experiments. On the molecular scale, such processes are a typical part of the overall dynamics reaching  from association over folding of biological macromolecules down to chemical reactions. The actual interest of theoreticians is in predicting equilibrium constants for end states and rates for transitions between them because that are the data obtainable from experiments. However, the real time scale is often orders of magnitude larger and prevents direct observation of processes during simulation. On the other hand, theories like transition state theory are at hand for calculating kinetic and equilibrium data from knowledge of the potential energy function without solving equations of motion of equivalent methods. The feasibility is due to an enormous reduction of the high dimensional configuration space to one or a few reaction coordinates and a minimum of features along a reaction path.\\
Definition of a reaction coordinate (RC) subject to a sliding constraint seems to be the most natural way for collecting data that allow full characterization of the reaction path, and was recently declared the "method of choice" after a comparison of related methods [1]. Actually, constrained ensembles, sometimes also referred to as "blue-moon" ensembles [2], were only considered and treated theoretically in the nineties. Currently, they are used in different types of simulations to generate and evaluate pathways of activated processes.  The large and still increasing number of applications is probably due to the increasing computing capacity. Numerous biomolecules in aqueous environment have meanwhile been studied by classical simulations with constraints, for instance the large systems containing the chaperonin GroEL [3,4] or the rotatory F-1-ATPase [5]. On the other hand, single chemical reaction steps are analyzed in much detail by demanding ab-initio calculations [6-8]. QM/MM hybrid calculations were performed to treat enzyme catalysis or reactions in solution [9].\\
In a classical picture, activated processes take place in a potential energy landscape $V(x_1 ... x_N)$ by starting from a local minimum A, climbing up to a saddle point called the activated state and finally ending in a second minimum B. A minimum-energy pathway (MEP) connecting A and B is a curve associated with a monotonically increasing reaction coordinate $r(x_1 ... x_N)$ and a free energy A(r) which has a maximum at the activated state situated at $r = r^{\not=}$ if free energy is dominated by energy. For the calculation of transition rates, knowledge of $A(r^{\not=})-A(r_A)$ is crucial, but difficult to obtain from simulations because the probability $P(r^{\not=})$ to find the system in the activated state is minimal along the pathway and extremely small in all interesting cases. Like often in the field, the occurrence of rare events hampers the statistical evaluation due to insufficient sampling if no particular measures are taken.  This is due to the relation \\
\begin{eqnarray}
A(r) =  - {k_B}T\ln Q(r) =  - {k_B}T\ln P(r) - {k_B}T\ln Q
\end{eqnarray}
between probability \textit{P(r)} and the partition function for a given value of \textit{r}, \textit{Q(r)}. \textit{Q} is the full partition function. We shall develop the theory for Helmholtz free energy \textit{A}, but the final formula will hold for Gibbs free energy \textit{G} as well when mean values are taken at constant pressure and temperature. The brief outline is indicating the essential problems that have to be solved when describing an activated process by means of molecular dynamics simulations: once that the states A and B are known one has (a) to determine pathways, (b) to assign reaction coordinates, and (c) to calculate free energy \textit{A(r)}. The order of the first two tasks is not mandatory. The focus of this review is on the third task, but we will also touch the other topics since they are all interrelated.\\
When probability cannot be evaluated directly from simulation data by histogram methods, there are essentially two possibilities to make sure sufficient sampling everywhere. The first is application of a so-called umbrella potential that restrains the system near the hypersurface $r(x_1 ... x_N) = r_{set}$ of the desired RC [10]. This widespread method called umbrella sampling [11] requires postprocessing of data in order to correct for the umbrella potential and has been developed further recently [12]. The second method tracing back to Carter et al. [13] is based on the idea of a conditional ensemble \textit{E} where the RC takes exactly a given value  $r(x_1 ... x_N) = r_{set}$, which would allow unrestricted sampling everywhere along the pathway. The corresponding partition function is [14]\\
\begin{eqnarray}
Q(r) = \int {\exp \left( { - \beta V\left( {{\bf{\hat q}};r} \right)} \right)} g{\left( {{\bf{\hat q}};r} \right)^{{\textstyle{1 \over 2}}}}{d^{N - 1}}q
\end{eqnarray}
i.e. among \textit{N} generalized coordinates $q_1 ... q_N$ one selects $r=q_N$ as RC. The symbols $(x_1 ... x_N)$ are reserved for Cartesian coordinates and $\beta = 1 / {{k_B}T}$.  The derivative of free energy is a mean force\\
\begin{eqnarray}
\frac{dA}{dr} =  - {k_B}T\frac{d}{dr}\ln Q(r)
 = {\left\langle {\frac{{\partial V}}{{\partial r}}} \right\rangle _E} - \frac{1}{2}{k_B}T{\left\langle {\frac{{\partial g}}{{\partial r}}} \right\rangle _E} \equiv  - {\left\langle f \right\rangle _E}
\end{eqnarray}
that is composed of a mean potential force and a second, entropic contribution originating from the mass-metric tensor determinant $g({{\bf{\hat q}};r})$. Although attractive from the theoretical point of view, this approach poses serious problems in numerical praxis. The conditional ensemble \textit{E} does not coincide with the constrained ensemble C that would be generated by applying the holonomic constraint $r(x_1 ... x_N) = r_{set}$ which implicitly also entails vanishing velocity 
$\dot{r}(x_1 ... x_N) = 0$. 
Therefore, it was early proposed without proof to compute the mean force from the constraint ensemble and to replace the derivative of the potential by the negative constraint force [15]. The crucial step forward was made soon afterwards in a rigorous analysis of the constrained case by M\"{u}lders et al. [16]. They could show that in this case no mass metric tensor contribution occurs and the mean force is exactly the negative constraint force, which opened a practicable way to compute numerically relevant mean forces and the free energy. At the same time it was clear that the full problem was not yet solved, and the solution would require considering the metric tensor effect. To be viable, the solution should also avoid the partial derivative with respect to the RC in (3). Note that its evaluation requires definition of really all coordinates $q_1 ... q_N$ which is often extremely difficult and practically impossible. Den Otter and Briels [17] showed that use of $g({{\bf{\hat q}};r})$ can be avoided by considering the Fixman determinant $z$ and the mean force can be written \\
\begin{eqnarray}
{\left\langle f \right\rangle _E} \equiv  - \frac{{dA}}{{dr}} =  - {\left\langle {{f^c}} \right\rangle _c} + {\Phi _{OB}}\left( z \right)
\end{eqnarray}
where $\Phi_{OB}(z)$ depends on and the gradient of \textit{z} that can be evaluated using Cartesian coordinates without regress to the generalized coordinates $q_1 ... q_N$. In a similar way Sprik and Ciccotti [18] were able to derive an equivalent expression with a formally different correction term $\Phi_{SC}(z)$. Thus, theory had reached a stage where it could be applied successfully to the numerical calculation of free energy profiles for interesting cases. Generalizations were made to allow for instantaneous instead of constraint force with a correction $\Phi_{DP}(z)$ [19], multiple constraints often applied in molecular dynamics simulations and more-dimensional reaction coordinates [19,20]. Later, theory was reconsidered in order to get a unified concise formulation of the confusing seemingly incompatible expressions $\Phi_{XY}(z)$ found earlier for the necessary correction term. In fact, a simple form was determined and proven to coincide with previous proposals. The theory part of this review will follow this derivation [21,22] which offers a relatively direct approach by starting from the constrained case.\\
The methods mentioned so far and compared in [1] can be subsumed as equilibrium methods based on equilibrium densities (histogram or umbrella sampling)or equilibrium mean constraint force. Non-equilibrium methods have been popular in some other fields of free energy calculation, but for the potential of mean force such an approach called 'dynamic umbrella sampling' was only recently designed by Hummer and Szabo, see [23]. \\
The constraint method is related to others from the wide field of free energy calculations [14,23], in particular to thermodynamic integration. The preceding considerations suggest a short-hand notation $H_c(\textbf{q,p};r)$ for the constrained Hamiltonian where \textit{r} is only a parameter. The mean force given by $\langle f \rangle_c = - \langle \delta H_c/\delta r\rangle_c$ determines free energy as the potential of mean force. This is a formulation one would expect when naively transferring the formalism of thermodynamic integration to the present problem replacing the $\lambda$-parameter by the RC. As a matter of fact, this is possible only for the constrained case. In general, a coordinate is more closely connected with dynamics than a  $\lambda$-parameter and needs a suitable treatment. Nevertheless, the simple relation $\partial A/\partial r = \langle \partial H_c/\partial r \rangle _c$ holds for some interesting reaction coordinates as will be proven below.
The review starts in section 2 with the search for pathways and properties of reaction coordinates. The choice of appropriate RCs and pathways are indispensable steps preceding free energy calculation. Section 3 outlines the essential parts of theory. Numerical applications from different fields and comparisons of methods are reviewed in the last section 4 followed by a summary.

\section{Pathways and reaction coordinates}
\label{sec:pathways}

Suitable pathways and reaction coordinates are the ingredients for full characterization of activated processes. Only equilibrium constants can in principle be determined from the end states of a process by calculating absolute free energy. The usual way is, however, the calculation of relative free energy along a reaction path.

\subsection{Search for pathways}
\label{sec:search_path}

The idea of stable states of reacting molecules and reaction pathways is influenced by pictures of a potential energy surface (PES) spanning over a two-dimensional area, i.e. by low-dimensional problems. Stable states are identified with local minima and pathways with minimum-energy paths  that connect them while saddle points on the way define activated complexes. The MEP is a one-dimensional curve $r(q_1 ... q_N)$ of all \textit{N} particle coordinates along which the probability density $P(q_1 ... q_{N-1};r) = P(\textbf{q};r)$ is maximal at fixed  \textit{r}. When it connects the minima of stable states, it is a reasonable reaction coordinate describing a productive reaction pathway (Figure 1A,C). There are numerous techniques for determining MEPs, for instance those described in [24] that were more and more refined in the course of time. The basic idea is definition of an initial trial path that is deformed to become a MEP. A more recent approach is transition path sampling [25] for reactive pathways designed to find paths connecting the end states by means of dynamics simulations. Note that these methods are reaction coordinate-free methods. Once that a path is known, one has a sequence of not necessarily equidistant points in configuration space that can be assigned values $r_1 ... r_n$ of an RC.\\
\begin{center}
\begin{figure}
\includegraphics[scale=0.62]{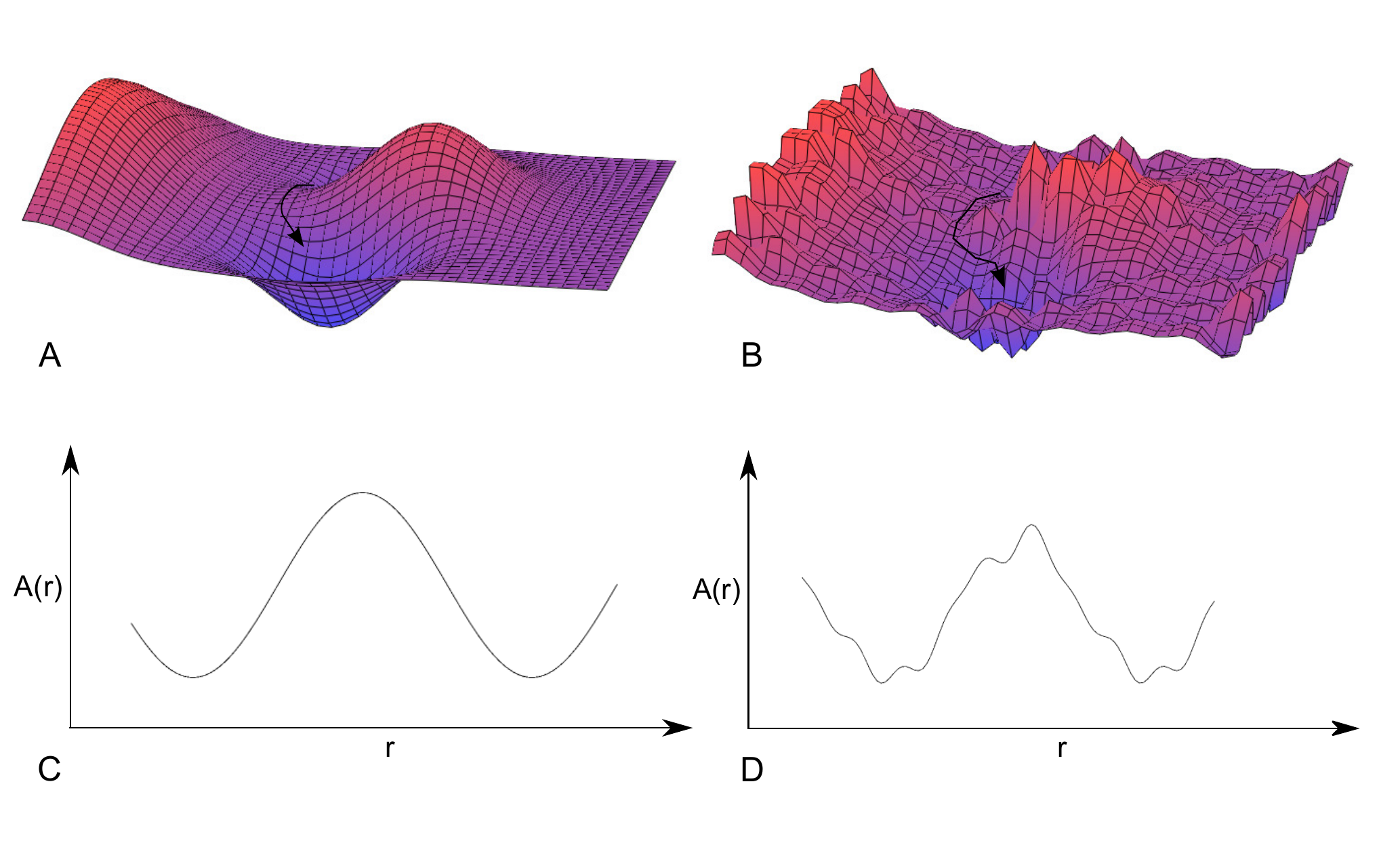}
\caption{Excamples of potential energy surfaces (top) and free energy profiles (bottom). Smooth surfaces and profiles are found at small molecules in vacuo (left). Rugged surfaces are typical of marcomolecules like proteins (right).}
\label{fig:taeler} 
\end{figure}
\end{center}
Macromolecules possess a much more complex PES than small molecules (Figure 1B,D). The glass transition in proteins, for instance, has proven experimentally the existence of a vast number of substates  separated by barriers which are permanently crossed at temperature above 200-250 K [26]. In simulations, this behavior was confirmed by the observation that about every 150 fs at 300 K the protein is crossing a barrier towards the next minimum in the PES [27]. The definition of a pathway as a MEP makes no more sense at finite temperature as multiple possible trajectories connect basins on a rugged energy landscape. The whole conformational space accessible from one point at fixed RC now belongs to the same pathway. Moreover, a bundle of separated pathways can be available like in small systems. Methods were designed to allow for these characteristics and to tackle the enormous costs of simulating biological macromolecules in their natural environment. The search for a path that connects the known folded with an unfolded conformation of a protein [28] based on the principle of least action is an example for the extension of ideas developed for small systems with new techniques. The methods mentioned so far aimed to determine first a pathway and then assign an RC for evaluation. At large molecules, it can make sense to start from a predefined, preliminary RC and then to drive the system along the RC towards the desired end state or away from a starting structure. This is done by application of a series of  constraints  \\
\begin{eqnarray}
r(t) = {r_i};\,\,\,\,\,\,\,\,\,\,\,\,\,i = 0,1,\,...\,m;\,\,\,\,\,\,\,\,\,\,\,\,\,{r_0} > {r_1} > ...{r_m}
\end{eqnarray}
with increasing (or decreasing) RC where the starting structure is situated at $r = r_0$ and the last structure at $r = r_m$ is a known or undefined structure that is to be determined that way. The constraint can also be time-dependent according to\\
\begin{eqnarray}
r(t) = {r_0} + \left( {{r_m} - {r_0}} \right)t/T\,\,\,\,\,\,\,\,\,\,\,\,\,\,\,\,\,\,\,\,0 < t < T
\end{eqnarray}
and increase (or decrease) the RC during a simulation of period \textit{T}. The first RC employed in this way for a macromolecule was the TMD coordinate defined as the rms distance of selected atoms from their position in a ‘target’ structure that will be discussed below. The series of RC values was generated by a time-dependent \textit{r(t)} in a MD simulation. It was employed to find a path for a secondary structure transition in insulin [29], later applied to an unfolding problem [30] in a similar way and has found many applications in large protein conformational transitions, for instance [31]. Driving an RC should not be confused with the ‘slow growth’ technique for free energy calculation. It serves merely for generating pathways independent of the method by which they are later characterized thermodynamically.

\subsection{Reaction coordinate}
\label{sec:reaction_coordinate}

The common idea of elementary chemical reactions or association/dissociation is suggesting to use a distance between nuclei as a RC [32,33].  At more complex reactions, the RC of single steps may not suffice to define a path and can be combined to a new RC as just mentioned [34]. For rotation [35] and isomerization [20], dihedral angles are appropriate RCs. The references represent only examples of a vast number of cases where RCs were employed to describe, simulate or fully evaluate reaction pathways of molecular activated processes. \\

In many cases, distance-like coordinates are a good choice for defining a preliminary RC even at complex situations. A very general form for \textit{n} particles involved is \\
\begin{eqnarray}
r({\bf{x}}) = {\left( {\sum\limits_{i = 1}^{3n} {{\mu _i}^2{{\left( {{x_i} - {y_i}} \right)}^2}} } \right)^{1/2}} = {\left( {\sum\limits_{j = 1}^n {{\mu _j}^2{{\left( {{{\bf{r}}_j} - {{\bf{s}}_j}} \right)}^2}} } \right)^{1/2}}
\end{eqnarray}
which is a mass-weighted rms distance between a configuration ${\bf{x}}(t) = (x_1 ... x_{3n}) = ({\bf{r}}_1 ... {\bf{r}}_n)$ and a fixed target or reference configuration ${\bf{y}} = (y_1 ... y_{3n}) = ({\bf{s}}_1 ... {\bf{s}}_n)$. The \textit{3n} Cartesian coordinates belong, for instance, to all or some selected atoms of a macromolecule. Mass-weighting of the coordinates as introduced by factors $\mu ^2_j = m_j /$ $m^*$ can be important as will be shown below. If, however, the masses are not too different and $m^*$ is chosen to be the mean mass $m^* = M/n$ of a system of total mass \textit{M}, then \textit{r} is approximately the geometrical distance of \textit{{\bf{x}}} and \textit{{\bf{y}}} in configurational space, and the rms distance between the configurations if $m^*$ is set equal to \textit{M}. This RC applies to different problems where it has different meaning:

\begin{description}
\item [(a)] \textit{Distance from a surface.} To describe adsorption of a molecule on a plane \textit{x = 0}, the first sum in (2.3) is restricted to $x_1$ which then denotes the center of mass distance of the molecule from the surface.
\item [ (b) ]	\textit{Two-particle distance.} With the restriction to $N=1$ , $r$ can describe the (mass-weighted) distance between two atoms or the centers of mass of two molecules at positions $x$ and $y$, respectively, which is interesting for association/dissociation reactions. When the atoms coincide with the first and last atom of a chain molecule, $r$ becomes the end-to-end distance similar to the RC of steered dynamics [36].
\item [(c)] \textit{Distance between conformers.} After superposition of the structure \textit{{\bf{x}}} of a given molecule  to the reference structure \textit{{\bf{y}}} by first translating and then rotating \textit{{\bf{x}}}, the resulting minimum distance measures the structural (rms) distance between the conformers. This sort of distance is used in targeted molecular dynamics simulations  (TMD) [29]
\item [(d)] \textit{Radius of gyration.} \textit{r} is the radius of gyration when all reference positions $r_j$ are replaced by the position of the center of mass, \textbf{R} and $m^* = M$, as then
\end{description}

\begin{eqnarray}
r({\bf{x}}) = {\left( {\sum\limits_j^{} {\frac{{{m_j}}}{M}{{\left( {{{\bf{r}}_j} - {\bf{R}}} \right)}^2}} } \right)^{1/2}}
\end{eqnarray}
A similar, but different RC is the dynamic distance [6] defined by\\
\begin{eqnarray}
r = {\left( {\frac{1}{{N_P^{}}}\sum\limits_{nops\,p = 1}^{{N_p}} {\frac{{{\mu _i}_j}}{{m^*}}{{({{\bf{r}}_i} - {{\bf{r}}_j})}^2}} } \right)^{{1 \mathord{\left/
 {\vphantom {1 2}} \right.
 \kern-\nulldelimiterspace} 2}}}
\end{eqnarray}
Here the sum runs over all $N_p$ non-overlapping pairs of selected atoms. The expression $\mu _{ij} = m_im_j/(m_i + m_j)$ is the reduced mass of the respective pair, and $m^*$ an arbitrary constant mass. This RC shares the favorable properties discussed in the following due to appropriate mass-weighting. With the restriction to \textit{n = 1, r} can describe the (mass-weighted) distance between two atoms or the centers of mass of two molecules at positions \textit{{\bf{x}}} and \textit{{\bf{y}}}, respectively, which is interesting for association/dissociation reactions. When the atoms coincide with the first and last atom of a chain molecule, \textit{r} becomes the end-to-end distance, i.e. the RC of steered dynamics [26].

\subsubsection{Mechanical properties}
\label{sec:mech_prop}

As we have seen, reaction coordinates are employed to drive a system from the start to the target state or to measure the constraint force, which is the topic of the next chapter. In any case, a force is applied to the system under consideration that can cause side effects like translation, rotation or the like.  There may be applications where such motions are in the focus of the interest. Normally, they represent undesired features that can be discarded by using suitably designed RCs. As an example for distance-like coordinates, we study the coordinate (8) $r(x_1 ... x_{3n}) = r({\bf{r}}_1 ... {\bf{r}}_n)$ fixed by the constraint\\
\begin{eqnarray}
\sigma \,(r;{r_0}) = r - {r_0} = {\left( {\sum\limits_j^{} {{\mu _j}^2{{\left( {{{\bf{r}}_j} - {{\bf{s}}_j}} \right)}^2}} } \right)^{1/2}} - {r_0} = 0
\end{eqnarray}
It results in constraint forces
\begin{eqnarray}
{{\bf{f}}_j}^c = {f^c}\frac{d}{{d{{\bf{r}}_j}}}\sigma ({\bf{x}}) = {f^c}\frac{d}{{d{{\bf{r}}_j}}}r({\bf{x}}) = {f^c}{\mu _j}^2\left( {{{\bf{r}}_j} - {{\bf{s}}_j}} \right)/r
\end{eqnarray}
The total constraint force
\begin{eqnarray}
{\bf{F}}_{tot}^c \equiv {\sum\limits_{} {{{\bf{f}}_j}} ^c} = {f^c}\sum\limits_{} {{\mu _j}^2\left( {{{\bf{r}}_j} - {{\bf{s}}_j}} \right)/r} 
\end{eqnarray}
is apparently proportional to the difference between the centers of mass of the momentary structure $({\bf{r}}_1 ... {\bf{r}}_n)$ and the target $({\bf{s}}_1 ... {\bf{s}}_n)$ as $\mu _j^2 = m_j / m^*$, and vanishes when they have been superimposed. A similar argument proves that also the total torque $M_{tot}^c = \sum r_j \times f_j^c$ vanishes after superimposing [29,35]. Note that mass-weighting is an important ingredient for these favorable properties which prevent rigid-body motions to disturb the molecular dynamics.\\

\subsubsection{Statistical mechanical properties}
\label{sec:stat_mech}

It is a well known fact that geometrical constraints can disturb the probability density function (PDF) of the unconstrained system in configurational space, $P(q_1 ... q_N)$.  If a constraint is imposed to $r = q_N$ it gives rise to a different PDF $P_c(q_1 ... 1_{N-1};r)$. However this is not the case for the above defined distance. As proven by Fixman [37] the probabilities are connected by
\begin{eqnarray}
{P_c}({q_1}\,...\,{q_{N - 1}};r) \propto \sqrt {\det ({\bf{H}})} P\left( {{q_1}\,...\,{q_{N - 1}}{q_N}} \right)
\end{eqnarray}
where the so-called Fixman determinant is\\
\begin{eqnarray}
z \equiv \det ({\bf{H}}) = \sum {\frac{1}{{{m_i}}}} {\left( {\frac{{\partial \;r}}{{\partial \;{x_i}}}} \right)^2}
\end{eqnarray}
For the distances (8) and (10) one easily derives $z = 1/m^*$, which is a constant. This proves that the PDF in configurational space is not changed by introducing the constraint \textit{r = const}. Already the basic formula (1) indicates that also free energy is not affected by the constraint if \textit{z = const}, which facilitates free energy calculation drastically.

\section{Theory}
\label{sec:Theory}

In the introduction, the basic ideas and problems on the way to a free energy profile were  already outlined. This section presents the essential steps of a rigorous derivation of the free energy profile.

\subsection{Probability and free energy}
\label{sec:prop_free}

As shown before, the concept of a reaction coordinate \textit{r} amounts to computing the probability \textit{P(r)} or the free energy \textit{A(r)} for one selected coordinate $r = q_N$ of N coordinates $q_1 ... q_N$ with velocities $v_i = \dot{q}_i$ and associated canonical momenta $p_i$. With the convention $\textbf{q} = (q_1 ... q_N)^T$ for coordinates and ${\bf{v}} = \dot{\textbf{q}}$ for velocities the Hamiltonian can be written as\\
\begin{eqnarray}
{H_{}}({\bf{q}},{\bf{p}}) = V({\bf{q}}) + {\textstyle{1 \over 2}}{{\bf{v}}^T}{\bf{Av}} = V({\bf{q}}) + {\textstyle{1 \over 2}}{{\bf{p}}^T}{{\bf{A}}^{ - 1}}{\bf{p}}
\end{eqnarray}
where \textit{V} denotes the potential energy, and $\bf{A}$ is the mass-metric tensor that contains the masses $m_i$ [37] and defines momenta by $\textbf{p} = {\bf{Av}}$. The corresponding partition function is (up to a constant factor)\\
\begin{eqnarray}
Q = \int {d{q^N}\,d{p^N}\exp ( - \beta \;H)} 
\end{eqnarray}
For a given value of the reaction coordinate, the  PDF is given by the reduced partition function \\
\begin{eqnarray}
Q(r) = \int {d{q^N}\,d{p^N}\exp ( - \beta \;H)} \,\delta (r - r') = \int {d{q^{N - 1}}\,d{p^{N - 1}}d{p_r}\exp ( - \beta \;H)} 
\end{eqnarray}
as \textit{P(r) $\equiv$ Q(r)/Q = exp (- $\beta$ A(r))/Q}. The associated free energy defined as \textit{A(r) = } $ - k_BT $ \textit{ln Q(r)} is a function of the RC that is often denoted as the free energy profile. By taking the derivative of (18) with respect to the RC one obtains a crucial relation between the mean force $\langle f \rangle$ and free energy (1),\\
\begin{eqnarray}
\left\langle f \right\rangle  \equiv  - \left\langle {\frac{{\partial H}}{{\partial r}}} \right\rangle  =  - \frac{{dA}}{{dr}}
\end{eqnarray}
Apparently, free energy is the potential of mean force (PMF). In practice, it is impossible to compute numerically \textit{P(r)} except for small systems where histogram methods can be applied. However, it is possible to introduce a constrained system satisfying the two conditions $q_N = r = const$ and $\dot{q}_N$ = 0. The restriction of velocity enables sampling for arbitrary periods and measuring, for instance, the mean constraint force which was early expected by van Gunsteren [14] to deliver free energy and proven later by M\"{u}lders et al. [16]. The constrained Hamiltonian $H_c$ satisfying both conditions is\\
\begin{eqnarray}
{H_c}({\bf{\hat q}},{\bf{\hat p}};r) = V({\bf{q}}) + {\textstyle{1 \over 2}}{{{\bf{\hat v}}}^T}{\bf{a\hat v}} = V({\bf{q}}) + {\textstyle{1 \over 2}}{{{\bf{\hat p}}}^T}{{\bf{a}}^{ - 1}}{\bf{\hat p}}
\end{eqnarray}
Here ${\bf{\hat q}} = (q_1 ... q_{N-1})^T$ contains only the first $N-1$ coordinates and the kinetic energy depends only on the restricted set of velocities ${\bf{\hat v}} = (\dot{q}_1 ... \dot{q}_{N-1})^T$ or  momenta ${\bf{\hat p}} = (\hat p_1 ... \hat p_{N-1})^T$ defined by ${\bf{\hat p}} = {\bf{a \hat v}}$. $\bf{a}$ is the mass-metric tensor of the constrained system. The free energy $A_c(r) = -k_BTlnQ_c(r)$ for this constrained system is given by its partition function \\
\begin{eqnarray}
{Q_c}(r) = \int {d{q^{N - 1}}\,d{{\hat p}^{N - 1}}\exp ( - \beta \;{H_c}({\bf{\hat q}},{\bf{\hat p}};r))} \,
\end{eqnarray}
We now focus on the free energy change $dA/dr$ in (1) that is obtained directly from $Q(r)$ according to \\
\begin{eqnarray}
\frac{{dA}}{{dr}} =  - {k_B}T\frac{d}{{dr}}\ln Q(r)
\end{eqnarray}
or, using the constrained quantities, as\\
\begin{eqnarray}
\frac{{dA}}{{dr}} = \frac{{d{A_c}}}{{dr}} - {k_B}T\frac{d}{{dr}}\ln \frac{{Q(r)}}{{{Q_c}(r)}}
\end{eqnarray}
It turns out that all of these quantities can be obtained in the constrained system. $-dA_c / dr = \langle f \rangle _c$, the mean force measured in the unnaturally constrained system, will be treated in the next section. After that, the correction term is calculated. The derivative $-dA / dr \equiv \langle f \rangle$ is the desired mean force which by numerical integration yields the free energy profile $A(r)$.

\subsection{Constrained case}
\label{sec:con_case}

Like above in (14) , the mean force is also in the constrained system given by \\
\begin{eqnarray}
 - d{A_c}/dr = {\left\langle f \right\rangle _c} = {\left\langle { - {{\partial {H_c}} \mathord{\left/
 {\vphantom {{d{H_c}} {dr}}} \right.
 \kern-\nulldelimiterspace} {\partial r}}} \right\rangle _c}
\end{eqnarray}
On the other hand, a mean constraint force $\langle f^c \rangle _c$ occurs  and can be measured numerically as a consequence of the applied constraint. The usual equations of motion in Cartesian coordinates are Lagrangian equations of the first kind like\\
\begin{eqnarray}
{m_i}{{\ddot x}_i} =  - \frac{{\partial V}}{{\partial {x_i}}} + \lambda \frac{{\partial r}}{{\partial {x_i}}}
\end{eqnarray}
where the Lagrange parameter is identical with the constraint force, $\lambda = f^c$ and results from the condition $r = r_{set} = const$. Anticipating the result of the following detailed analysis by M\"{u}lders et al. [6], the mean force turns out to be the negative mean constraint force\\
\begin{eqnarray}
{\left\langle f \right\rangle _c} = {\left\langle { - {{d{H_c}} \mathord{\left/
 {\vphantom {{d{H_c}} {dr}}} \right.
 \kern-\nulldelimiterspace} {dr}}} \right\rangle _c} =  - {\left\langle {f_{}^c} \right\rangle _c}
\end{eqnarray}
The mean force in the constrained system is the negative mean constraint force. The derivation of (26) is very subtle since it requires switching from Lagrangian to Hamiltonian formalism. One has to consider in more detail the mass-metric tensor \textit{\textbf{A}} which is defined by its matrix elements\\
\begin{eqnarray}
{A_{kl}} = \sum\limits_i {{m_i}\frac{{\partial {x_i}}}{{\partial {q_k}}}\frac{{\partial {x_i}}}{{\partial {q_l}}}} 
\end{eqnarray}
and contains mass-weighted derivatives of a set of Cartesian coordinates with respect to the generalized coordinates $\left( {{q_1}\,...\,{q_{N - 1}}{q_N}} \right) = \left( {{{{\bf{\hat q}}}^T}{q_N}} \right) = \left( {{{{\bf{\hat q}}}^T}r} \right)$. \textit{\textbf{A}} and its inverse can be written as block matrices,\\
\begin{eqnarray}
{\bf{A}} = \left( {\begin{array}{*{20}{c}}
{\bf{a}}&{\bf{b}}\\
{{{\bf{b}}^{\bf{T}}}}&c
\end{array}} \right){\rm{    }}\,\,\,\,\,\,\,\,\,\,\,\,\,\,\,\,\,\,\,\,\,\,\quad {{\bf{A}}^{ - 1}} = \left( {\begin{array}{*{20}{c}}
{\bf{x}}&{\bf{y}}\\
{{{\bf{y}}^{\bf{T}}}}&z
\end{array}} \right)
\end{eqnarray}
where \textbf{a} is the $(N-1) \times (N-1)$ matrix used above. Then the free Lagrangian is $L = {\textstyle{1 \over 2}}{\left( {{{{\bf{\dot{\hat q}}}}^T}\dot r} \right)^T}{\bf{A}}\left( {{{{\bf{\dot{\hat q}}}}^T}\dot r} \right)-V\left({{{{\bf{\hat q}}}^T}r} \right)$. In contrast, the constrained Lagrangian with only \textit{N-1} dynamical coordinates reads\\
\begin{eqnarray}
{L_c}\left( {{\bf{\hat q}},{\bf{\dot{\hat q}}};r} \right) = {\textstyle{1 \over 2}}{{{\bf{\dot{\hat q}}}}^T}{\bf{a\dot{\hat q}}} - V\left( {{{{\bf{\hat q}}}^T};r} \right)
\end{eqnarray}
 and the corresponding Hamiltonian\\
\begin{eqnarray}
{H_c}\left( {{\bf{\hat q}},{\bf{\hat p}};r} \right) = {\textstyle{1 \over 2}}{{{\bf{\hat p}}}^T}{\bf{a\hat p}} + V\left( {{{{\bf{\hat q}}}^T};r} \right)
\end{eqnarray}
with canonical momenta ${\bf{\hat p}} = {{\partial {L_c}} \mathord{\left/ {\vphantom {{\partial {L_c}} \partial }} \right. \kern-\nulldelimiterspace} \partial }{\bf{\dot{\hat q}}}$ contain the RC only as a parameter.  Using the identity ${{\partial {{\bf{a}}^{ - 1}}} \mathord{\left/ {\vphantom {{\partial {{\bf{a}}^{ - 1}}} {\partial r}}} \right. \kern-\nulldelimiterspace} {\partial r}} =  - {{\bf{a}}^{ - 1}}\left( {{{\partial {\bf{a}}} \mathord{\left/ {\vphantom {{\partial {\bf{a}}} {\partial r}}} \right. \kern-\nulldelimiterspace} {\partial r}}} \right){{\bf{a}}^{ - 1}}$, one finds the relation\\
\begin{eqnarray}
\frac{{\partial {H_c}\left( {{\bf{\hat q}},{\bf{\hat p}};r} \right)}}{{\partial r}} =  - \frac{{\partial {L_c}\left( {{\bf{\hat q}},{\bf{\dot{\hat q}}};r} \right)}}{{\partial r}}
\end{eqnarray}
On the other hand, Lagrangian equations of the first kind can be formulated using ${L^{(1)}} \equiv L + \lambda r$ where $\lambda$ is the constraint force subject to the condition $r = const$. $L^{(1)}$ is usually expressed by all \textit{N} Cartesian coordinates $(x_1 ... x_N)$ in numerical applications. For this argument, however, the above generalized coordinates $(q_1 ... q_{N-1}q_N)$ are chosen. We then evaluate the derivates at \textit{r = const} and $\dot{r} = 0$ which yields\\
\begin{eqnarray}
\begin{array}{l}
\cfrac{{\partial {H_c}\left( {{\bf{\hat q}},{\bf{\hat p}};r} \right)}}{{\partial r}} = \lambda  - \cfrac{\partial {L^{(1)}}}{\partial r}\\
 = \lambda  - \cfrac{d}{{dt}}\cfrac{{\partial {L^{(1)}}}}{{\partial \dot r}}\\
 = \lambda  - \cfrac{d}{{dt}}\left[ {{\bf{b}}({\bf{\hat q}};r){\bf{\dot{\hat q}}}} \right]{\,_r}
\end{array}
\end{eqnarray}
Here, we first employed (31), then inserted the general Lagrangian equations, and finally made use of the mass-metric tensor (28). In order to finish the proof, one has to remember that the ensemble average of a time derivative must vanish and the Lagrange parameter $\lambda$ is identical to the constraint force. Thus, it holds ${\left\langle {{{d{H_c}}\mathord{\left/ {\vphantom {{d{H_c}} {dr}}} \right. \kern-\nulldelimiterspace} {dr}}} \right\rangle _c} = \left\langle \lambda  \right\rangle _c  = {\left\langle {f_{}^c} \right\rangle _c}$ which finally proves eq. (26).

\subsection{Correction for the unconstrained case}
\label{sec:corr_case}

For calculating the correction term in (23) the partition function $Q(r)$ is transformed so that it can be related to the one of the constrained case, $Q_c(r)$. We switch to the Hamiltonian formalism and make use of canonical momenta. A decomposition of the momentum vector, ${\bf{p}} = {({\tilde p_1}...{\tilde p_{N - 1}}{p_r})^T} = {({{\bf{\tilde p}}^T},{p_r})^T}$ is chosen in accordance with the above introduced block structure of the mass-metric tensor (3.13). This allows one to decompose the kinetic energy [17] as\\
\begin{eqnarray}
\begin{array}{l}
K = {\cfrac{1}{2}}{{\bf{p}}^T}{{\bf{A}}^{ - 1}}{\bf{p}}\\
\quad   = {\cfrac{1}{2}}{{{\bf{\tilde p}}}^T}{{\bf{A}}^{ - 1}}{\bf{\tilde p}} + {\textstyle{1 \over 2}}{({p_r} + {z^{ - 1}}{{\bf{y}}^T}{\bf{\tilde p}})^T}z({p_r} + {z^{ - 1}}{{\bf{y}}^T}{\bf{\tilde p}})
\end{array}
\end{eqnarray}
Inserting \textit{K} in the partition function makes it possible to perform the integral over $p_r$ and to obtain a very simple result,\\
\begin{eqnarray}
\begin{array}{l}
Q(r) \\ = \int {{{(dq\,d\tilde p})^{N - 1}}d{p_r}\exp \left( { - \beta \left( {V + {\textstyle{1 \over 2}}{{{\bf{\tilde p}}}^T}{{\bf{a}}^{ - 1}}{\bf{\tilde p}} + {\textstyle{1 \over 2}}{{({p_r} + {z^{ - 1}}{{\bf{y}}^T}{\bf{\tilde p}})}^T}z({p_r} + {z^{ - 1}}{{\bf{y}}^T}{\bf{\tilde p}})} \right)} \right)}  \\
   = \int {d{q^{N - 1}}\,d{{\tilde p}^{N - 1}}\exp \left( { - \beta \left( {V({\bf{q}}) + {\textstyle{1 \over 2}}{{{\bf{\tilde p}}}^T}{{\bf{a}}^{ - 1}}{\bf{\tilde p}}} \right)} \right)} \;{z^{ - {\textstyle{1 \over 2}}}}({\bf{q}}) \cdot const \\
    = \int {d{q^{N - 1}}\,d{{\tilde p}^{N - 1}}\exp \left( { - \beta \,{H_c}({\bf{\hat q}},{\bf{\tilde p}};r)} \right)} \;{z^{ - {\textstyle{1 \over 2}}}}({\bf{q}}) \cdot const \\
 = {\left\langle {{z^{ - {\textstyle{1 \over 2}}}}} \right\rangle _c}{Q_c}(r) \cdot const
\end{array}
\end{eqnarray}
Note that $H_c (\bf{\hat q, \tilde p}; r)$ is the Hamiltonian of the constrained system defined in (30), and \textit{Q(r)} the corresponding partition function which is inserted here to transform the integral into a thermal average. One finds for the ratio of the  partition functions
\begin{eqnarray}
Q/{Q_c} = const \cdot \,{\left\langle {{z^{ - 1/2}}} \right\rangle _c}
\end{eqnarray}
where the quantity \textit{z} is the well known Fixman determinant of the coordinate transformation used the average of which is a function of the RC. We can now rewrite the differential form of the free energy (23) as\\
\begin{eqnarray}
\frac{{dA}}{{dr}} = {\left\langle {{\lambda _r}} \right\rangle _c} - {k_B}T\frac{d}{{dr}}\ln {\left\langle {{z^{ - 1/2}}} \right\rangle _c}
\end{eqnarray}
Integration reveals a simple form which is easily evaluated numerically,\\
\begin{eqnarray}
A(r) = \int\limits_{}^{} {{{\left\langle {{\lambda _r}} \right\rangle }_c}} dr - {k_B}T\ln {\left\langle {{z^{ - 1/2}}} \right\rangle _c}
\end{eqnarray}
So far, the analysis was made for the case of only one constraint coordinate, the RC \textit{r}. The straightforward extension to multiple constraints is found in other works[20,22,33].\\ 

According to the fundamental relation (1), the probability for finding the system between $r $and $r + dr$ is
\begin{eqnarray}
P(r)dr = const \cdot exp (-A(r)/k_B T)dr
\end{eqnarray}
Thus, the free energy profile $A(r)$ reflects the equilibrium probability distribution in terms of barriers and wells, and enables application, for instance, of transition state theory. However, the interpretation may sometimes be difficult because a Jacobian determinant can be contained in $P(r)$ and distort the intuitively expected profile.\\ 

As an example, consider the distance in configuration space defined in (8). It is known from applications of TMD that the profile always increases dramatically at decreasing distance. When the constraint acts on $d$ cartesian coordinates, $d-1$ degrees of freedom are left and the Jabobian is  $r^{d-1}$. One might be more interested then in the probability $g(r)$ for finding the system in a volume element $dV = r^{d-1} dr$ at distance $r$. Obviously $P(r)dr=g(r)r^{d-1}dr$. The corresponding free energy profile hence reads
\begin{eqnarray}
A_g(r)=(1-d)ln r + A(r)
\end{eqnarray}
This profile can be calculated analytically [35] and turns out to be constant if the potential energy landscape is flat. Otherwise it represents the energetics of a transition in a reasonable and comprehensive way.

\section{Applications}
\label{sec:appcl}

Although the focus of this review is on methodology, a brief glimpse on the very different applications of the theory is to illustrate the scope and benefit of simulations that are fully evaluated with respect to pathways and free energy. \\

A few examples of free energy calculations on small molecules were already given above for demonstrating the use of simple RCs. The pioneering papers did apply the method to an ion pair in solvent [32] and dihedral angles  [20,35]. The photoinduced proton transfer in the Watson-Crick GC base pair was studied using the dynamic distance [8] in Car-Parinello simulations. The simulations revealed the sequence of elementary proton transfer steps, which is a typical result of application of RCs that comprise many particle coordinates like (8) and (10), and explained the spontaneous repair after irradiation by the shape of  the free energy profile. Car-Parinello simulations with constraint are also the background of numerous other free energy calculations of chemical reactions [7].\\

The discussion of PES and pathways has adumbrated the problems of applications of the full method to macromolecules. It is certainly prohibitive to solve the docking problem with pathway search as discussed above, but undocking of ligands and dissociation of dimers of proteins are feasible. For instance, the dissociation of two superoxide dismutase molecules was studied in much detail and characterized by a free energy profile [33]. The dissociation of a phenyl molecule from a insulin complex was investigated in a similar way [38]. Most applications of constraints to activated processes, however, do without thermodynamic evaluation. They are restricted to the characterization of pathways with respect to details which can be verified experimentally, ranging from rather local [39] to large scale conformational transitions [4,5,40], folding/unfolding [31,41] and many other applications of the TMD coordinate (8).\\

It has been known for a long time that numerical free energy calculations do not converge when performed with a slow-growth technique, but require application of a windows scheme [42]. Slow growth of the RC (7) is, however, suited to generate pathways. However, the approximate free energy profile \textit{A(r(t))} calculated ‘on the fly’ from the momentary constraint force can indicate a sequence of events like bond ruptures and is hence a useful tool to explore details of a pathway. In this regard, it is superior to an energy profile \textit{E(t)} as free energy tends to be much less noisy.\\

A window scheme means performing a constraint simulation at \textit{n} subsequent values of the RC (6) for getting reliable mean force values (37) from the constraint force, i.e. the Lagrangian parameter. The force tends to fluctuate with considerable amplitudes[33,38], which requires control of convergence. The free energy profile is obtained then by numerical integration.\\

A very clear and detailed comparison of the constraint method with umbrella sampling was published by Trzesniak et al. [1] who applied a few variants of both methods to the association of two methane molecules in water with the same simulation period. Of course, results will in general depend on the problem and the choice of the RC. The constraint method needs computation of the Fixman determinant in cases that deviate from those discussed above in section 2. On the other hand, there is no need to find to optimum restraint potential. In summary, the authors arrive at the result that both methods yield comparable results, the constraint method being the best choice. \\

\section{Summary}
\label{sec:summ}

Although successful in many applications, the concept of a reaction coordinate is still under debate for principle reasons as well as for numerical problems arising eventually. In favor of the method, one finds the argument that fixing a single coordinate (the RC) during a simulation run at finite temperature leaves a system practically full freedom to adapt the constraint (or restraint) by relaxation and equilibration in all coordinates but one. The path is a multidimensional entity that allows flexibility despite numerous small barriers most of which are probably due to transient hydrogen bonds when proteins in aqueous solution are considered. Convergence of the mean force can take a long time and must be monitored carefully.\\

Nevertheless, other equally broad pathways may exist which are not detected this way, but possibly by repeating computation with different starting conditions. This is not a shortcoming of the RC approach, but a difficulty inherent to large systems. It poses a practical problem that can only be tackled by increasing numerical efforts. There seems to be no way to solve the problem mathematically. A second problem is the definition of an appropriate RC even for a small system. The optimum RC would measure the distance traveled along the underlying MEP, otherwise most of the transition takes place in the hyperplane orthogonal to the progress in  the RC. A reliable indication for this is a sudden jump in the constraint force which is eventually observed, for instance, at accompanying proton transfer [36]and can be avoided by swapping the RC. It would also possible to compare the traveled distance in configuration space with the one in the RC in order to monitor this kind of behavior. \\ 

Constraints have proven to be a useful tool for simulating activated processes, in particular for the final calculation of free energy. Therefore they have been applied to an impressive number of problems ranging from local rotations to protein dissociation. For generating pathways, constraint methods are particularly suited when large molecules or complexes are studied, while for small systems also other methods are available. It is in principle possible to replace a constraint everywhere by a restraint potential, and there is a tendency to do so because simulation packages allow to add a potential, but do not provide a simple way to implement a constraint. Constraints possess the advantage that postprocessing is superfluous and performed best in a realistic, stringent comparison. We have also shown that a metric-tensor correction is not needed in practice for many reaction coordinates. If necessary, it is easily calculated from a simple formula derived recently.\\

\section{Acknowledgment}
\label{sec:ack}

The author thanks Martina Bamberg and Katrin Augustinowski for their help at the preparation of the manuscript.

\end{document}